\documentclass[12pt]{iopart}
\usepackage{graphicx}
\usepackage{bm}
\begin{document}
\title{Phase separation and the effect of quenched disorder in $Pr_{0.5}Sr_{0.5}MnO_3$}
\author{A K Pramanik and A Banerjee} 
\address{UGC-DAE Consortium for Scientific Research, University Campus, Khandwa Road
Indore-452017, M.P, India.}
\ead{alok@csr.ernet.in}
\date{\today}

\begin{abstract}
The nature of phase separation in $Pr_{0.5}Sr_{0.5}MnO_3$ has been probed by linear as well as nonlinear magnetic susceptibilities and resistivity measurements across the 2nd order paramagnetic to ferromagnetic transition ($T_C$) and 1st order ferromagnetic to antiferromagnetic transition ($T_N$). We found that the ferromagnetic (metallic) clusters, which form with the onset of long-range order in the system at $T_C$, continuously decrease their size with the decrease in temperature and coexist with non-ferromagnetic (insulating) clusters. These non-ferromagnetic clusters are identified to be antiferromagnetic. Significantly, it is shown that they do not arise because of the superheating effect of the lower temperature 1st order transition. Thus reveals unique phase coexistence in a manganite around half-doping encompassing two long-range order transitions. Both the ferromagnetic and antiferromagnetic clusters form at $T_C$ and persist much below $T_N$. Substitution of quenched disorder (Ga) at Mn-site promotes antiferromagnetism at the cost of ferromagnetism without adding any magnetic interaction or introducing any significant lattice distortion. Moreover, increase in disorder decreases the ferromagnetic cluster size and with 7.5\% Ga substitution clusters size reduces to the single domain limit. Yet, all the samples show significant short-range ferromagnetic interaction much above $T_C$. Resistivity measurements also reveal the novel phase coexistence identified from the magnetic measurements. It is significant that, increase in disorder up to 7.5\% increases the resistivity of the low temperature antiferromagnetic phase by about four orders. 
\end{abstract}


\maketitle
\section {Introduction}
Physical properties and their variation with external stimuli or internal disorder in half doped perovskite manganites with generic formula $R_{0.5}A_{0.5}MnO_{3}$ (where R and A respectively stand for trivalent rare earth and divalent alkaline earth elements) are topics of significant current interest \cite{dago, tokura}. Coexistence of contrasting phases, namely, ferromagnetic (FM) - metallic (M) and antiferromagnetic (AF) - insulating (I) is a common occurrence in manganites around half doping \cite{dago}. This phase coexistence or phase separation (PS) is considered to be responsible factor for observed functional properties as well as argued to be main impetus for \textit{bi-critical} phase competition \cite{tokura}. However, the origin of PS and the nature of coexisting phases remain a matter of debate \cite{little, alokjpcm}. It is widely considered that FM-M and AF-I phases have similar energies in manganites around half-doping as a result of which small perturbations, compared to thermal energy, can cause colossal changes \cite{dago, tokura, hotta}. This proximity in energies of two contrasting phases remains despite considerable change in the nature of spin, charge and orbital ordering of the ground state with the variation in ionic radii of R/A atoms, which affects the Mn-O-Mn bonds and the one electron bandwidth \cite{dago, tokura}. The decrease in average radii of R/A atoms brings about decrease in bandwidth and the ground state of the system changes from so called A-type to CE-type AF structure. This indicates that the nature of PS and the effect of quenched disorder may not remain identical as spin, charge and orbital ordering changes across the bandwidth.
       
It has been shown that `finite-size' clusters with large uncompensated spins are created when quenched disorder is introduced in the Mn-site of narrow bandwidth $Pr_{0.5}Ca_{0.5}MnO_3$ having CE-type AF and charge ordered-insulating ground state \cite{sunil}. Electronic phase separation gives rise to interesting effects within these clusters having some similarities with other transition metal oxides like cuprates and nickelates \cite{tranq1, tranq2}. Quenched disorder in the form of magnetic ions (a few \% of Cr and Co) in $Nd_{0.5}Ca_{0.5}MnO_3$ and $Pr_{0.5}Ca_{0.5}MnO_3$ respectively, gives rise to coexisting AF-I and FM-M phases with many unusual physical properties at low temperature \cite{kimura, mahen}. Interestingly, recent studies have shown that the ground state of $Pr_{0.5}Ca_{0.5}MnO_3$ changes to FM-M even with minimal substitution of non-magnetic disorder (2.5\% Al) at Mn-site, without introducing any significant structural distortion \cite{alokjpcm, banerjee}. Significantly, the conductivity of $Pr_{0.5}Ca_{0.5}MnO_3$ increases when such disorder is introduced in the Mn-O-Mn network \cite{suniljpcm}. Apart from chemical substitution, internal disorder in various forms also lead to remarkable effects. For example, anisotropic stress in $Nd_{0.5}Sr_{0.5}MnO_3$, having larger bandwidth than $Pr_{0.5}Ca_{0.5}MnO_3$ , has significantly changed the orbital ordering (OO) and other physical properties \cite{nsmo}. In addition, $Pr_{0.5}Sr_{0.5}MnO_3$, which is an intermediate bandwidth system with A-type AF-I ground state, shows strain induced dimensionality crossover triggering a localization-delocalization transition \cite{uozu}. In the classic work of Imry and Ma \cite{imry}, it was argued that random quenched disorder arising from lattice defects, dislocations or chemical substitution can destabilize the long range order system favoring formation of `finite-size' clusters. Appearance of coexisting clusters and their size regulation by disorder in manganites is also shown from simulation work by Moreo \textit{et al.} \cite{moreo}. In another study, Burgy \textit{et al.} proposed chemical disorder driven inhomogeneous states across the first order transition \cite{burgy}. Recently, the fragility of both A-type and CE-type AF-I ground state against quenched disorder is shown from computer simulation \cite{alvarez}. Following these experimental and theoretical developments, it becomes imperative to investigate seriously the nature of PS and the effect of disorder in different half doped manganites.
                    
In this context, $Pr_{0.5}Sr_{0.5}MnO_3$ (PSMO) is very attracting system considering the A-type AF ground state where FM layers are coupled antiferromagnetically exhibiting quasi two dimensional (2D) metallic behavior \cite{kawano,hejt}. Moreover, this composition is at the phase boundary of FM and AF ground states \cite{kajimoto} thus, making it quite vulnerable towards tendency for PS, which has been observed at low temperature from the $^{55}Mn$ NMR study \cite{allodi}. Further, PSMO has a second order paramagnetic (PM)-I to FM-M transition around 270 K and a first order FM-M to AF-I transition around 140 K \cite{tomioka}. Imry and Wortis \cite{wortis} had predicted the rounding of first order transition due to quenched random disorder and beyond certain percentage the nature of transition is expected to change from first to second order. Recently, it is shown that the transformation kinetics of the first order transition in PSMO is hindered, resulting in tunable coexisting fraction of kinetically arrested high-T (FM) phase with equilibrium AF-I phase at low temperature  \cite{alokjpcm}. Hence, substitution of quenched disorder in this system will reveal the effect on both the phase transitions as well as on the nature of PS in the same system. However, introducing quenched disorder in manganite in the form of chemical substitutions remains a challenging task, as introduction of dissimilar ion(s) at R/A or Mn-site can lead to change in structure, which modifies the original system completely. Apart from this, introduction of magnetic elements will also modify the basic system by introducing additional magnetic interactions. 

We report here a detail study of PS in PSMO and the effect of substitutional disorder (Ga) at the Mn-site. Substitution of Ga neither add any magnetic interaction nor introduce any significant lattice distortion since $Ga^{3+}$ being $d^{10}$ element has zero orbital or spin moment and has matching ionic radii with the existing $Mn^{3+}$ \cite{suniljpcm}. Earlier study on the same substitution in PSMO has mainly focused on magneto-transport properties and shown that high-T FM phase is suppressed and low-T AF phase is enhanced due to the quenched disorder \cite{maignanZ}. Thus disorder promotes the AF interactions at the cost of FM in PSMO, which is contrary to the observation in $Pr_{0.5}Ca_{0.5}MnO_3$ for the similar kind of substitution \cite{banerjee,alokjpcm}. The significant findings of present study are summarized below. We show drastic decrease in FM transition temperature ($T_C$) and increase in AF transition ($T_N$) with Ga substitution (upto 7.5\%) without any significant change in structure. The change in $T_C$ with Ga follows simple mean-field variation indicating that there is no unwanted magnetic interaction is added to the system by the quenched disorder (Sec. 3.1). The coexisting FM and non-FM clusters arise at high temperature just below $T_C$ and persists much below $T_N$ for all the samples. Size of the FM clusters reduces with the increase in quenched disorder (Sec. 3.2). The FM phase fraction changes continuously as the temperature is reduced from $T_C$ and is tracked using second-order susceptibility which directly probes the variation in the spontaneous magnetization (Sec. 3.3). An attempt is made to identify the nature of the coexisting non-FM phase in the FM regime and is found to be AF in nature. \emph{This significant observation implies that both the FM and AF phases form at $T_C$} (Sec. 3.4). It is interesting to note that though the measured magnetic moment at 2K matches well with the fully spin aligned value, the moment found from the fitting to Curie-Weiss Law above $T_C$ gives much higher value indicating the existence of FM short-range order above $T_C$ (Sec. 3.5). The intriguing nature of the coexisting phases in PSMO is substantiated through resistivity measurements which also shows drastic effect of the quenched disorder (Sec. 3.6). Significance of this study in the context of PS and the effect of quenched disorder in the intermediate bandwidth system around half-doping is provided in concluding section (Sec. 4).

\begin{table}
\caption{\label{tab:table 1} Structural and fitting parameters determined from the Rietveld profile refinement of the powder XRD patterns for the series Pr${_{0.5}}$Sr${_{0.5}}$Mn${_{1-x}}$Ga${_x}$O${_{3}}$. Here O1 refers to the apical oxygen of the perovskites and O2 is the equatorial oxygen which lies in the plane of the perovskite layer. The average Mn valence state of Mn was determined by iodometric titration. The extreme right column shows the percentage change ($\Delta\%$) of the parameters between the end members of the series (i.e. between x=0 and 7.5\%) .}
\begin{indented}
\item[]\begin{tabular}{cccccc}\\
\hline
Ga (x) &$0\%$ &$2.5\%$ &$5.0\%$ &$7.5\%$ &$\Delta\%$\\
\hline
a (\AA) &5.4027 &5.4068 &5.4044 &5.4046 &+0.07\\
c (\AA) &7.7814 &7.7820 &7.7817 &7.7791 &-0.03\\
V (\AA${^3}$) &227.13 &227.49 &227.24 &227.23 &+0.04\\
Mn-O1 (\AA)  &1.9454 &1.9455 &1.9451 &1.9448 &-0.03\\
Mn-O1-Mn  &180\raisebox{1ex}{\scriptsize o} &180\raisebox{1ex}{\scriptsize o} &180\raisebox{1ex}{\scriptsize o} &180\raisebox{1ex}{\scriptsize o} &0.0\\
Mn-O2 (\AA)  &1.9255 &1.9274 &1.9274 &1.9304 &+0.25\\
Mn-O2-Mn  &165.5\raisebox{1ex}{\scriptsize o} &165.32\raisebox{1ex}{\scriptsize o} &164.89\raisebox{1ex}{\scriptsize o} &163.65\raisebox{1ex}{\scriptsize o} &-1.1\\
Mn$^{3+}$\%  &51.74 &50.39 &47.51 &45.04 &-\\
Mn$^{4+}$\% &48.22 &47.1 &47.48 &47.45 &-\\
Mn Valance (Av) &3.4819 &3.4827 &3.4994 &3.5126 &+0.8\\
\hline
\end{tabular}
\end{indented}
\end{table}  

\section{Experimental Details}
Polycrystalline samples of $Pr_{0.5}Sr_{0.5}Mn_{1-x}Ga_{x}O_{3}$ series with x = 0.0, 0.025, 0.05 and 0.075 have been prepared by the standard solid state ceramic route using $Pr_6O_{11}$, $SrCO_3$, $MnO_2$ and $Ga_2O_3$ with purity more than $99.99\%$ and final sintering temperature was used $1500^{o}C$ for 36 hours. The x-ray diffraction (XRD) measurements were done with Rigaku Dmax 300 diffractometer with $CuK_\alpha$ radiation at room temperature. All the samples were found to be in the single phase and XRD pattern were analyzed by the Rietveld profile refinement programme by Young \textit{et al.} \cite{young}. To estimate the $Mn^{3+}/Mn^{4+}$ ratio, Iodometric redox titration has been done using sodium thiosulphate ($Na_2S_2O_3, 5H_2O$) and potassium iodide (KI). The average atomic concentration in system has been found out through energy dispersive analysis of x-ray (EDAX) attached to transmission electron microscope (TECNAI G2-20FEI). Low field AC susceptibility measurements were performed with the home-made AC-susceptometer \cite{ashnarsi}. DC magnetizations were measured with a home-made vibrating sample magnetometer (VSM) \cite{krsnarsi} and Quantum Design 14 Tesla VSM (PPMS). DC resistivity measurements were done with the standard four-probe method.   
    
Room temperature XRD pattern has been fitted and found to be in tetragonal \textit{I4/mcm} space group. All the Rietveld refinement parameters and the titration results have been given in Table 1. The \textit{goodness of the fit}, which is defined as the ratio of $R_{wp}$/$R_{exp}$ has been found to be around 1.4 for the whole series \cite{young}. It has been found that there is no major structural distortion due to doping, as seen in Table 1. Titration results show average Mn valence enhances with the doping. This indicates that substituted Ga has preferentially replaced $Mn^{3+}$ and it is also expected from ionic size matching ($Mn^{3+}$ = 0.65 \AA\ and $Ga^{3+}$ = 0.62 \AA). The average concentration of chemical constituents found through EDAX matches with the nominal concentration within experimental accuracy which is $\pm$ 0.5\% for the lowest Ga doped compound. 

\section{Results and discussions}
\subsection{Variation of T$_C$ and T$_N$ with quenched disorder}  
The real and imaginary part of the first order ac susceptibility ($\chi_1^R$ and $\chi_1^I$ respectively) measured for this series in 0.5 Oe ac field and frequency of 131 Hz is shown in the Fig. 1a and 1b respectively. Parent compound (x = 0) shows two magnetic phase transitions with lowering temperature (PM to FM and FM to AF). For the doped samples, PM to FM transition temperature ($T_C$) decreases and FM to AF transition temperature ($T_N$) increases systematically. Transition temperatures are calculated from the maximum change in $d\chi/dT$ with temperature and given in the Table 2. For x = 0 compound, $T_C$ matches well with the reported single crystal value \cite{tomioka}. Imaginary part of the ac susceptibility which signifies the magnetic loss also shows sharp changes around $T_C$ and $T_N$. 

\begin{table}
\caption{\label{tab:table 2}Transition temperatures have been calculated from the differential change of the ac susceptibility with the temperature for the series $Pr_{0.5}Sr_{0.5}Mn_{1-x}Ga_{x}O_{3}$. Last row shows measured demagnetization factor (N) at temperature $\approx 0.99T_C$.}
\begin{indented}
\item[]\begin{tabular}{ccccc}\\
\hline
Ga (x) &$0 \%$ &$2.5 \%$ &$5.0 \%$ &$7.5 \%$\\
\hline
$T_C$(K) &269.2  &236.41 &201.17 &174.26\\
$T_N$(K) &124.72 &133.59 &158.85 &-\\
$N$ &0.308 &0.533 &0.657 &-\\
\hline
\end{tabular}
\end{indented}
\end{table}

The large systematic change in $T_C$ ($\approx$ 100 K) without considerable structural distortion in present series is intriguing as the substituted (nonmagnetic) disorder does not add any magnetic interaction in the system.  We have calculated the tolerance factor ($\tau$) for the series and found very small change in $\tau$ ($\approx$ 0.03 \%) between the end compositions (0\% and 7.5\%).  To get similar change in $T_C$ by substitution at Ln-site in half doped compound $Ln_{0.5-x}Sr_{0.5}MnO_{3}$, large change in $\tau$ (around  0.4 - 0.9 \%)  is required \cite{damayTc}. Thus, the reduction of $T_C$ in $Pr_{0.5}Sr_{0.5}Mn_{1-x}Ga_{x}O_{3}$ arises from site dilution of the magnetic lattice and we attempted to explain this by mean field theory (MFT), considering only nearest neighbor interactions. According to MFT, $T_C$ can be expressed as \cite{kittel}:
\begin{equation}
	T_C = \frac{2S(S+1)zJ}{3k_B} 
\end{equation}
Where S is the average spin per magnetic ion, $k_B$ is the Boltzmann constant, z is the number of nearest neighbor magnetic atoms and J is the exchange integral. For manganite, Eq. 1 has to be modified considering possible magnetic interactions viz.  i) Super Exchange (SE) $Mn^{3+}-O^{-2}-Mn^{3+}$, $Mn^{4+}-O^{-2}-Mn^{4+}$ and ii) Double Exchange (DE) $Mn^{3+}-O^{-2}-Mn^{4+}$.  Considering these three interactions, variation of $T_C$ with Al substitution at the Mn-site in $LaMnO_{3+\delta}$ was explained \cite{krishnajpcm}.  However, for the present case we consider only an average magnetic interaction and if the Ga substitution (x) is random in the magnetic lattice, then it scales with the number of nearest neighbour (z) of Eq. 1. Thus a simple site dilution by Ga will result in a linear variation of  $T_C$ with x. Interestingly, Fig. 2 shows this linear variation of the experimentally found $T_C$ with Ga substitution (x) for the present series. This linear scaling of $T_C$ indicates that i) no significant complication is introduced due to site dilution or the average exchange integral is not modified and ii) substitution is random. It is interesting that this simple mean field approach has led to the qualitative understanding of the effect of disorder on modification of $T_C$ for such a complicated system. Similar analysis for large bandwidth manganite, $La_{0.7}Sr_{0.3}Mn_{1-x}M^\prime_xO_3$ ($M^\prime$ = Al, Ti) was used to explain linear variation of $T_C$ with the amount of Al as well as Ti substitutions \cite{nam}. Further, it was shown from the extrapolation of the linear variation, that $T_C$ vanishes only when $Mn^{3+}-O^{-2}-Mn^{4+}$ DE interaction vanishes by the selective substitution of either $Mn^{4+}$ or $Mn^{3+}$ by Ti or Al respectively. This was attributed to the dominance of DE interaction for this large bandwidth system. However, for the present Pr${_{0.5}}$Sr${_{0.5}}$Mn${_{1-x}}$Ga${_x}$O${_{3}}$ series, such extrapolation (from Fig. 2) leads to vanishing of  $T_C$ at x $\approx$ 0.21, when there are substantial amount of both $Mn^{4+}$ and $Mn^{3+}$ are present. This indicates the reduction in the strength of DE interaction with the decrease in bandwidth.

More drastic effect of quenched disorder is there on the $T_N$, which appears to be counterintuitive, and certainly opposite to general antiferromagnetic systems or even the 2-dimentional antiferromagnetic manganite \cite{sunil2d}. It is clear from Table-2 that $T_N$ increases monotonically with substitution. However, the variation of $T_N$ does not follow linear scaling with quenched disorder (x). Earlier study has shown that substitution of trivalent (In, Ga) and tetravalent (Sn, Ti) nonmagnetic elements at Mn site in PSMO increases and decreases $T_N$ respectively \cite{maignanZ}. Hence, respective change in Mn ionic concentration and minute structural modification arising from disorder can modify $T_N$ in opposite way. As seen in Table 1, Mn-O bond length in basal (O2) and apical (O1) plane has very small positive and negative changes respectively with disorder in this series. This can make FM state unstable and assist this A-type AF state (where FM interaction is realized along the basal plane and these planes are antiferromagnetically coupled along the apical direction) to set in at higher temperature, hence increasing $T_N$ with disorder. As mentioned earlier, substitution of Ga in the Mn-site of PSMO is one of the simplest route to introduce quenched disorder in the magnetic lattice of this system. However, the present experimental observations appear to contradict the theoretical proposal of fragility of A-type AF state against quenched disorder \cite{alvarez}. 

\subsection{Phase separation and the effect of quenched disorder on it}
We have probed the phase separation from the thermal hysteresis (TH) in ac-susceptibility. Fig. 3a shows $\chi_1^R$ measured in 0.2 Oe field and frequency of 131 Hz for the parent compound (x = 0) in both heating and cooling cycles. It is evident that $\chi_1^R$ shows finite (TH) which starts immediately below $T_C$ and persists much below $T_N$. The amount of TH [$\chi_1^R$(cooling) - $\chi_1^R$(heating)] as measured in 9 Oe ac-field and 131 Hz frequency is given as a function of temperature in Fig. 3b for the present series. In general, TH in ac-$\chi$ has been shown to be the generic feature of first-order phase transition (FM-AF) \cite{manekar}. But in addition to a peak around $T_N$, Fig. 3b shows that the TH starts immediately below the respective $T_C$s and persists through the FM phase for all the samples. This indicates the inhomogeneous nature of the FM state where finite size FM clusters coexist with non-FM one. Higher value of $\chi_1^R$ in the cooling run indicates that cooling run contains more FM fraction than the heating run. It is significant that the TH appears just below $T_C$ or with the onset of long range ordering in the system and persists below $T_N$. Moreover, similar TH is also observed in variation of spontaneous magnetization measured through higher order (even order) susceptibility as well as in resistivity and will be discussed in later sections.  \emph{This new type of PS across two long-range order magnetic transitions in the same sample is rather uncommon in literature. }

To understand the nature of the inhomogeneous FM state we have studied the change in magnetic anisotropy as a function of both temperature and composition. Ac-susceptibility shows sharp peak in low fields immediately below $T_C$, as shown in Fig. 1a. Enhancement of this peak height with decrease in field can be observed for x=0 sample in Fig. 3a for 0.2 Oe compared to that shown in Fig. 1a for 0.5 Oe.  Such peak is commonly known as \textit{Hopkinson's peak} and mainly originates due to the rapid increase in anisotropy immediately below $T_C$, in particular when it exceeds the applied ac field \cite{hein}. When both shape as well as magnetocrystalline anisotropies are present, the intrinsic magnetic susceptibility ($\chi_{int}$) is related to the measured low field susceptibility ($\chi_{mes}$) in the following way \cite{kaul}:                    
\begin{equation}
	\chi_{int}^{-1}(T) = \chi_{mes}^{-1}(T) - 4\pi N(T)
\end{equation}

Where N(T) = $N_d$ + $N_K(T)$, $N_d$ is called demagnetization factor which depends on the sample shape anisotropy and gives demagnetization field $H_d$ = $4\pi N_dM$  \cite{cullity}. The magnetocrystalline anisotropy is taken care by  $N_K$ and for polycrystalline FM sample, $N_d$ is the dominant factor to anisotropy. For a FM, $\chi_{int}$ diverges at $T_C$ in the absence of external magnetic field and $N_d$ depends only on the shape or dimension of the sample. Hence, the measured susceptibility in the ferromagnetic region of a homogeneous FM is expected to remain almost constant as a function of temperature (as shown in Ref. \cite{km}). However, significant temperature dependence in susceptibility within FM phase (Fig. 1a) indicates variation of magnetic anisotropy arising from the temperature dependence in the $N_d$ for the inhomogeneous FM state. Following the protocol given in Ref. \cite{kaul}, we have calculated $N_d$ at different temperatures in FM regime from low field dc magnetization data. We have also calculated coercive field ($H_C$)  from the hysteresis loops at various temperatures. Fig. 4 shows that $N_d$ increases with lowering temperature for x = 0 compound and exceeds the value (0.495) which was estimated from the dimension of the bulk sample \cite{osborn}. This clearly indicates that the $N_d$ for an inhomogeneous FM is not given by the sample dimension but by the dimension of the `finite-size' FM clusters within it. Moreover, its variation with temperature follows the variation in the magnetic anisotropy resulting in concomitant increase in $H_C$ as shown in Fig. 4. Thus we infer from Fig. 4, that the FM clusters in this system spontaneously change their dimensions even well within the FM state right from their formation, and corroborate the conclusion drawn from the observed thermal hysteresis in Fig. 3. It may be argued that, as we decrease the temperature and approach $T_N$, FM cluster size decreases resulting in increase in surface anisotropy contributing substantially to the total magnetic anisotropy of the system.

To understand the effect of quenched disorder on the inhomogeneous FM state, we have measured the variation of  $N_d$ and $H_C$ with temperature for other members of the series and found qualitatively similar behaviour (not shown) as shown in Fig. 4.  Table 2 shows the value of $N_d$  at $\approx$ 0.99 $T_C$s of respective samples, with x = 0.0, 0.025 and 0.05. It is clear that $N_d$ increases with increase in quenched disorder (x) though the bulk dimension of all the samples are roughly same. It also implies, from the argument given above, that increase in quenched disorder reduces the size of the FM clusters. Such variation in `finite-size' cluster with disorder qualitatively agrees with the earlier studies on cuprate \cite{cho}. In the present system, FM clusters form immediately below $T_C$ and coexist with non-FM clusters.  Size of these FM clusters reduce with decrease in temperature or increase in quenched disorder.   

We give further evidence of the decrease in FM cluster size with increase in quenched disorder from the field dependence of ac-$\chi$.  Fig. 5a shows $\chi_1^R$ for parent compound in the measuring ac-field range of 0.5 to 8 Oe. It is clear that the \textit{Hopkinson's peak} immediately below $T_C$,  though observed in 0.5 Oe, reduces and $\chi_1^R$ increase with increase in field. As mentioned earlier, \textit{Hopkinson's peak} which is an outcome of the rapid increase in anisotropy compared to the measurement field, is expected to decrease with increase in field. Higher field will help in overcoming higher pinning potentials and the domain walls will move further within the multi-domain FM clusters, giving rise to concomitant increase in $\chi_1^R$. On the contrary, such field dependence is absent for the x= 0.075 sample (Fig. 5b) indicating that the cluster size is so small that it cannot support multiple domains. We plot the difference between the $\chi_1^R$ measured in 8 Oe and 0.5 Oe at 0.99 $T_C$ of the respective samples as a function of disorder (x) in Fig. 5c. The observed decrease in field dependence indicates that the decrease in FM cluster size with increase in disorder (x) actually reduces the number of domains within a cluster and approaches the single-domain limit for x=0.075, where no field dependence is expected. Thus the decrease in cluster size will reduce the extent of the domain wall motion as the field is increased. Consequently, the differences between the magnetic losses for 8 Oe and 0.5 Oe i.e. the difference between the corresponding $\chi_1^I$ will also reduce, as shown in the Fig. 5c.
 
\subsection{Temperature variation of FM cluster size probed through non-linear susceptibility}
We have utilized non-linear ac-susceptibility to probe the phase-separated state in this series; through the effective study of variation of spontaneous magnetization as a function of temperature as well as quenched disorder. Non-linear magnetic susceptibilities are important experimental tools which are used to unravel intricacies about the magnetic states in different systems \cite{sunil, ashna, sunil2}. However, their magnitude being couple of orders smaller than the linear part, serious efforts are involved to measure them. In general, magnetization (m) can be expanded in terms of magnetic field (h) as
\begin{equation}
 m = m_0 + \chi_1h + \chi_2h^2 + \chi_3h^3 + \chi_4h^4 + ...	
\end{equation}
Where $m_0$ is the spontaneous magnetization, $\chi_1$ ($\approx \partial m/\partial h$) is linear and $\chi_2$, $\chi_3$, $\chi_4$, etc. are non-linear susceptibilities. The even order susceptibilities ($\chi_2$, $\chi_4$, ...) arise for the systems which hold no inversion symmetry for m with respect to applied field i.e m(h) $\neq$ -m(-h). In another way, presence of symmetry-breaking field is required for the experimental observation of $\chi_2$ in ac susceptibility measurement. The origin of such symmetry breaking field can be either a superimposed external dc field or internal field in a ferromagnet arising from spontaneous magnetization. For this reason, the second order susceptibility ($\chi_2$ $\approx \partial^2m/\partial h^2$) has been probed to study the presence of spontaneous magnetization in different compounds \cite{akm,ranganathan}. For ferromagnet, as the temperature is decreased through $T_C$, spontaneous magnetization appears giving rise to the symmetry breaking field and resulting asymmetry in magnetization can be defined as $\Delta$m = m(h) - (-m(-h)). Across the $T_C$, sharp increase in spontaneous magnetization will result in a sharp increase in $\Delta$m. The rate of this increase will slow down as it is cooled further below $T_C$ and may approach saturation far below $T_C$. Consequently, $\chi_2$ which is proportional to $\partial \Delta m/\partial h$ will show a sharp negative peak around $T_C$, similar to a peak expected for internal $\chi_1$ ($\approx \partial m/\partial h$). However, this peak in $\chi_2$ is expected to be negative since the internal field crated by spontaneous magnetization is opposite to the direction to the external ac-field. Moreover, $\chi_2$ shows sharp feature whenever there is sharp change in internal field and finite value as long as there is variation in the internal field or spontaneous magnetization.

We have measured $\chi_2$ in the absence of external dc magnetic field while heating as well as cooling the samples. Fig. 6 shows $\chi_2$ for the series measured in 9 Oe ac field and fundamental frequency of 131 Hz. While cooling from the higher temperature, $\chi_2$ for the samples x = 0, 0.025 and 0.05 appears with the onset of FM and shows sharp negative peak around $T_C$. Below $T_C$ for all the samples we found a finite value of $\chi_2$ in FM state whose magnitude decreases with decrease in temperature. On further cooling, around $T_N$, $\chi_2$ again shows comparatively small negative peak. Both the peaks around $T_C$ and $T_N$ indicate sudden change in internal field. In addition, the small peak around $T_N$ shows considerable thermal hysteresis and shifts in peak position (insets of Fig. 6). This confirms a broad first order FM-AF transition in the form of supercooling and superheating of FM/AF phases. More significantly, there is a considerable difference in $\chi_2$ between the heating and cooling runs in FM regime and around $T_C$. This clearly indicates difference in FM phase fraction in heating and cooling run. Though such thermal hysteresis in $\chi_2$ can be justified for a broad first order FM to AF transition but the same around the second order PM to FM transition and also within the FM state is noteworthy. Thus the coexisting phase fractions varies immediately with the onset of long range ordering. The presence of $\chi_2$ for x = 0.075 compound in wide temperature range shows conclusively the existence of ferromagnetism in this compound. However, this FM phase is rather inhomogeneous indicated by the broadness of the peak over a wide temperature range. This finding is quite remarkable as susceptibility/magnetization behavior of this compound (see Figs. 5b) resembles spin glass like where $\chi_2$ cannot exist. Moreover, magnetization of this compound looks similar to that of x = 0.06 compound of the same series in Ref. \cite{maignanZ}, where it is concluded that FM vanishes within this amount of substitution. But we show in this study that, long range FM in PSMO exists even at higher concentration of nonmagnetic substitution. \emph{Thus, observed thermal hysteresis in $\chi_2$ conclusively shows that coexisting FM phase fraction changes as a function of temperature right from the formation around $T_C$ but continues to exist much below $T_N$.}
 
\subsection{Phase inhomogeneity and identification of high-T non-FM phase}
Now we attempt to identify the nature of the coexisting non-FM phase found immediately below $T_C$. Probable nature of non-FM phase could be the high-T PM or low-T AF phase. Intuitively this phase may be considered to be AF since in this system both FM and AF phases are considered to have similar energies. It is interesting to note that our recent study on the parent compound has clearly shown that first-order field-temperature induced FM-AF transition process is kinetically arrested resulting in persistence of high-T FM phase fraction down to the lowest temperature (5K), much below the closure of the hysteresis of the first-order process \cite{alokjpcm}. We have measured dc-magnetization of the parent compound in 517 Oe following different protocol. Fig. 7a shows considerable amount of TH in FM regime between field cooled cooling (FCC) and field cooled warming (FCW) magnetization, indicating that the coexisting phase fractions change during cooling and heating cycles. This measurement field (517 Oe) being much above the coercive field discards the possibility that TH arise due to the effects of local anisotropy. In addition, the same measurement has been repeated on the powder sample (after crushing the pellet) and we got TH similar to the bulk (not shown) which rule out any possibility arising form other factors like strain. Again, we have measured magnetization in 517 Oe while cooling in zero field, following a rigorous protocol similar to Ref. \cite{freitas}. In this protocol, we have started cooling the sample in zero field from above room temperature. At each measurement temperature, we isothermally apply 517 Oe field and measure the magnetization. After that, we isothermally reduce the field to zero, cool the sample to the next lower measurement temperature and repeat the above procedure for the successive lower temperatures. Magnetization measured in this protocol is shown as M(517, 0) in Fig. 7a. If the coexisting phase is PM (having the same structural symmetry with FM) then with an application of field, FM clusters will grow freely irrespective of the way the field is applied and M(517, 0) will match with FCC data. But the coexisting AF phase, structurally different from FM phase, will not grow freely as field is applied after cooling in zero field due to the energy barrier between the AF and FM phases. Consequently M(517, 0) will be less than FCC data. It may be noted that in PSMO, PM and FM phase are having same structure but there is structural change around FM-AF transition \cite{damay}.  Fig. 7a clearly shows M(517,0) departs from FCC below $T_C$ and gives lesser value than the M(FCC) throughout the measuring temperature. This may be considered as strong evidence that the coexisting non-FM phase is AF and not PM. To check whether application of field at each measurement temperature for M(517, 0) modifies the corresponding phase fraction irreversibly, we have measured M(virgin) by the following protocol. Each time we cooled the sample from the PM-state to target temperature in zero field, isothermally applied 517 Oe to measure magnetization M(virgin) for that temperature and repeat the complete procedure for all other temperatures . As evident in Fig. 7a, there is no difference between M(virgin) and M(517,0).
 
The above experiment strongly suggests that the AF phase also form around $T_C$ and coexist with the FM phase in the ferromagnetic region. In this region, an energy barrier separates the FM and AF phase and the later has higher energy than the former. The higher energy AF phase will be in metastable state which has been confirmed by the time dependent magnetization. Magnetization has been measured as a function of time (t) after cooling the sample in zero field from PM phase to the target temperature and subsequently applying a measuring field. Normalized magnetization (M(t)/M(0)) as a function of t is shown in Fig. 7b. A continuous increase in magnetization is observed after application of 100 Oe at 225 K and 150 K (within FM regime) and the same has also been observed after application of 5 kOe at 225 K. This suggests that the coexisting metastable phase (AF) tries to overcome the energy barrier to achieve the low energy FM state. The presence of AF phase again checked by collecting FCC and FCW data at 517 Oe but cooling was done down to only 170 K which is well within the FM phase and much above $T_N$. Fig. 7c shows a TH indicating presence of AF phase. \emph{This is a significant observation since it suggests that presence of AF phase at high temperature is not due to superheating phenomenon because the sample has not approached the $T_N$ while cooling. This also clearly indicates that both the FM and AF phases form with the onset of long-range order at $T_C$.} Neutron diffraction or other microscopic experimental tools could be used to directly confirm this significant observation. However, it may not be trivial to identify such small changes in the coexisting phase fractions unambiguously.
      
\subsection{Evidence of short range FM interaction above $T_C$}
Measured magnetization in 1 Tesla field shows Curie-Weiss behaviour, M/H = $\chi=C/(T-\theta_P)$ only from temperatures which are much higher than the respective $T_C$s (Fig. 8). Effective Bohr magneton ($\mu_{eff}$) per formula unit (f.u.) calculated from the fitting to Curie-Weiss law are significantly larger than the expected spin only value ($\mu_{eff} = g\sqrt{S(S+1)}$) and are given in Table 3. This indicates that the short-range FM interactions exist much above $T_C$. Existence of short range FM interaction much above $T_C$ was proposed for $La_{0.67}Ca_{0.33}MnO_3$ compound where magnetic clusters were detected from small angle neutron scattering by De Teresa \textit{et al.} \cite{teresanature}. Such kind of magnetic clusters in PM region have been shown for single crystal PSMO \cite{cao} and other manganites in both single crystal \cite{volkov} as well as in polycrystalline sample \cite{lu} and thought to be an intrinsic property of manganites \cite{dago}. It may be noted here that the field induced FM state at 2K gives almost fully aligned spin moment at 14 Tesla. The measured magnetic moment ($\mu$) per f.u. at 2K in 14 Tesla as well as the expected moment per f.u. calculated for spin only value from the respective $Mn^{3+}$ and $Mn^{4+}$ content is given in Table 3. The measured and calculated moment for spin only value match reasonably well for all the samples.

\begin{table}
\caption{\label{tab:table 3}Measured $\mu_{eff}$ determined from fitting to Curie-Weiss law, calculated  $\mu_{eff}$ considering spin only value, measured and calculated saturation moment ($\mu$) values determined from the magnetization data for the series Pr${_{0.5}}$Sr${_{0.5}}$Mn${_{1-x}}$Ga${_x}$O${_{3}}$.}  
\begin{indented}
\item[]\begin{tabular}{cccccc}\\
\hline
Ga (x) &$0 \%$ &$2.5 \%$ &$5.0 \%$ &$7.5 \%$\\
\hline
Measured $\mu_{eff}$ $(\mu_B/f.u.)$ &5.532 &5.479 &5.549 &5.449\\
Expected $\mu_{eff}$ $(\mu_B/f.u.)$ &4.396 &4.286 &4.161 &4.038\\
Expected $\mu$ $(\mu_B/f.u.)$  &3.516 &3.428 &3.325 &3.225\\
Measured $\mu$ $(\mu_B/f.u.)$ &3.508  &3.281 &3.261 &3.176\\
\hline
\end{tabular}
\end{indented}
\end{table}

\subsection{Phase coexistence and the resistivity measurement}
Resistivity measurement also substantiates the remarkable phase coexistence shown from the magnetic measurements. Fig. 9a shows resistivity for x = 0 compound measured while heating and cooling. A metal to insulator transition (MIT) has been observed around 270 K in agreement with the single crystal measurement \cite{tomioka}. As the FM-AF transition is approached while cooling, resistivity shows steep increase accompanied by a huge thermal hysteresis, which is characteristics of first order phase transition. Minor hysteresis loops (MHLs) are shown to be a useful method to confirm as well as study the disorder broadened first order phase transition \cite{manekar,kanwaljeet}. Following similar protocol we have recorded MHLs by measuring resistivity while cooling, to different temperatures (113, 104, 95 and 90 K) and then heating from those points to all the way above $T_N$ (Fig. 9b). MHLs are also recorded in heating cycle (not shown) confirming disorder broadened first order transition and related variation of coexisting metallic and insulating phase fractions. However, it is noteworthy that TH in resistivity remains much above $T_N$ (it remains distinct up to 225 K well within FM region) as shown in the inset of Fig. 9a. \emph{It is rather significant that resistivity being a percolative process, such TH well within the FM region confirms that fraction of FM (metallic) clusters changes right from their formation at higher temperature as also concluded from the magnetic measurements.} 

Fig. 10 shows resistivity for all the samples. No clear MIT was observed in the Ga substituted samples and resistivity increases with the substitution at all temperatures. However, the signature of FM to AF transition is clear from the kink in resistivity appearing around the $T_N$ of the respective samples with an exception for x = 0.075.  The effect of quenched disorder is drastic and shows an increase by about 0.5 $\times 10^4$ times in resistivity at 92 K, well within AF phase, with only 7.5\% substitutional disorder. Though increase in resistivity with quenched disorder is observed in another half doped bilayer manganite \cite{sunil2d} but the observed colossal increase in the present series is rather intriguing since here also the resistivity is governed by the electrical conduction in the FM layers of the A-type AF structure. It may be mentioned here, that in an earlier study decrease in resistivity has been observed with disorder for the CE type AF state \cite{suniljpcm}. In view of these and also in the context of recent theoretical developments \cite{alvarez}, detailed investigation of the resistivity behaviour of the present series needs to be undertaken.           
 
\section{Conclusion}
In summary, we have studied the structural, magnetic and transport properties of half doped $Pr_{0.5}Sr_{0.5}MnO_3$ and the effect of quenched disorder (Ga substitution) in the magnetic lattice without introducing any additional magnetic interactions or significant structural distortion. Substitution of Ga in the Mn-site has drastic but opposite effects on the FM and AF transitions. Increase in substitutional disorder decreases $T_C$  but increases $T_N$ and points toward reduction in the strength of double-exchange interaction in this intermediate bandwidth system. The FM state is found to be inhomogeneous and the FM cluster size decreases with decrease in temperature or increase in quenched disorder. The electronic phase separation gives rise to thermal hysteresis in the size of this clusters right from their formation at $T_C$ which is evident from the TH in the susceptibility or spontaneous magnetization. This system shows a novel phase coexistence, \emph{both FM and AF clusters form with the onset of long-range order in the system at $T_C$ and persist down to the lowest temperature.} Moreover, the short range FM interaction exists much above $T_C$. Resistivity also shows the signature of the variation of coexisting phases and substantiates the conclusion drawn from the magnetic measurements. At lower temperature (in AF state), resistivity shows orders of magnitude increase with quenched disorder. The observed novel phase coexistence needs further attention and required to be probed through microscopic tools and supporting theoretical work.

\section{Acknowledgment}
We express our sincere thanks to Dr. P. Chaddah for many helpful discussions regarding phase coexistence. We acknowledge Dr. N. P. Lalla for XRD and EDAX measurements. We thank Mr. Kranti Kumar and Mr. K. Mukherjee for the help during the measurements. DST, Government of India is acknowledged for funding 14 Tesla VSM. AKP also acknowledges CSIR, India for financial assistances.

\begin{figure*}
	\centering
		\includegraphics[width=7cm]{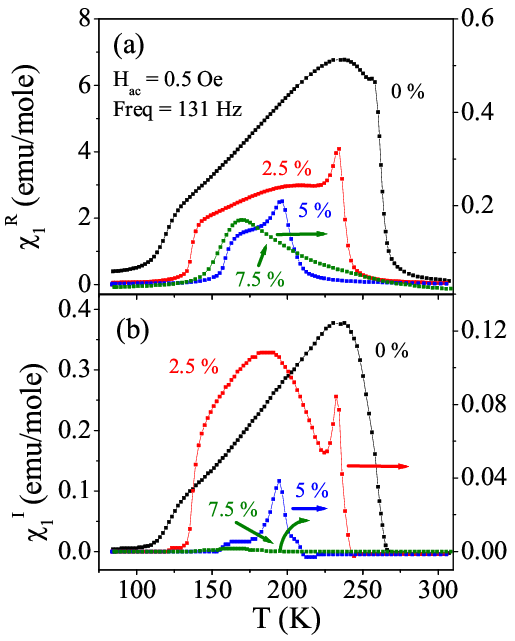}
		\caption{ (a) Real and (b) Imaginary parts of first order ac susceptibility ($\chi_1^R$ and $\chi_1^I$)) measured in an 0.5 Oe ac field and 131 Hz frequency have been plotted as a function of temperature for the 
series $Pr_{0.5}Sr_{0.5}Mn_{1-x}Ga_xO_3$.}
	\label{fig:fig1}
\end{figure*}

\begin{figure*}
	\centering
		\includegraphics[width=7cm]{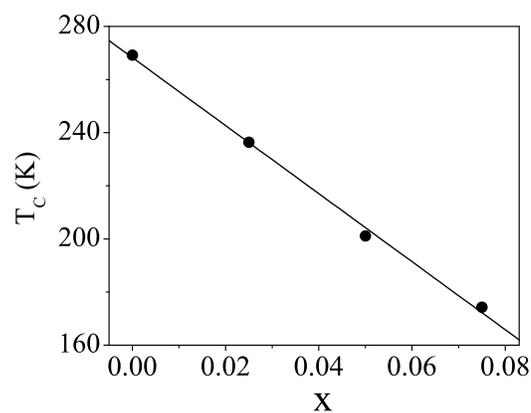}
		\caption{Variation of $T_C$ has been plotted as a function of x for the $Pr_{0.5}Sr_{0.5}Mn_{1-x}Ga_xO_3$ series. This figure shows linear decrease in $T_C$ with increase in disorder.}
	\label{fig:fig2}
\end{figure*}    

\begin{figure*}
	\centering
		\includegraphics[width=7cm]{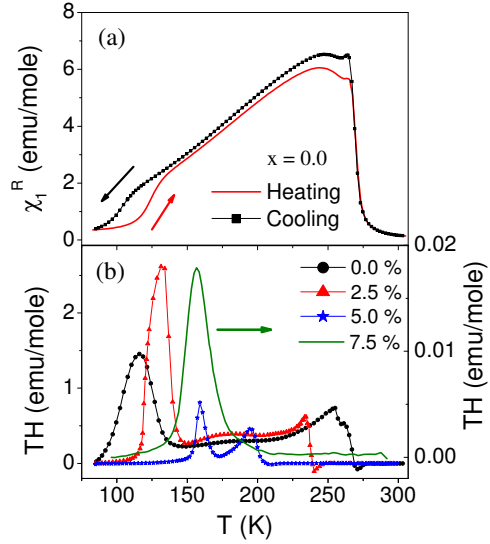}
		\caption{(a) Real part of first order ac susceptibility ($\chi_1^R$) measured in 0.2 Oe ac field and 131 Hz during heating and cooling for x = 0 compound. (b) Temperature variation of amount of thermal hysteresis (defined in the text) is shown for the first order susceptibility which has been measured in 9 Oe and 131 Hz for the series $Pr_{0.5}Sr_{0.5}Mn_{1-x}Ga_xO_3$.}
	\label{fig:fig3}
\end{figure*}

\begin{figure*}
	\centering
		\includegraphics[width=8cm]{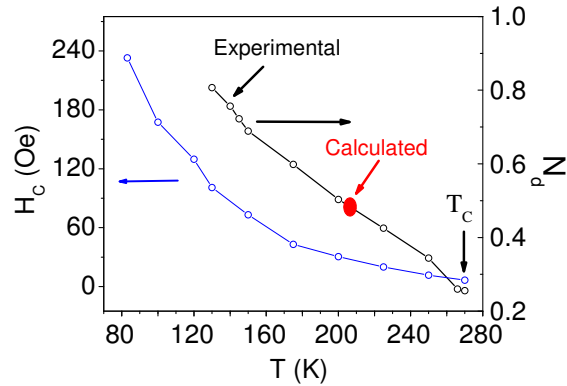}
	\caption{Left axis shows temperature variation of coercive force $H_C$ and right axis shows experimentally measured demagnetization factor (N$_d$) in the FM regime as a function of temperature for x = 0 compound. The filled symbol is the calculated N$_d$ for the same compound.}
	\label{fig:Fig4}
\end{figure*}

\begin{figure*}
	\centering
		\includegraphics[width=7cm]{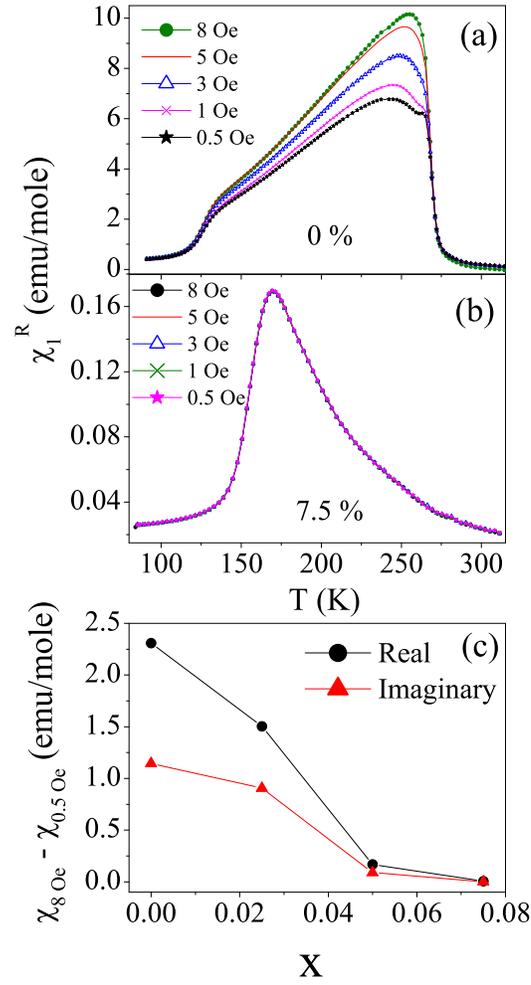}
		\caption{ (a) Temperature variation of real part of ac-susceptibility ($\chi_1^R$) measured in different ac field and 131 Hz frequency for x = 0 compound. (b) The same field dependence has been given for x = 0.075 compound. (c) Difference of $\chi$ between 8 Oe and 0.5 Oe fields as measured at 0.99T$_C$ for real and imaginary parts have been plotted as a function of composition (x).}
	\label{fig:fig5}
\end{figure*}  

\begin{figure*}
	\centering
		\includegraphics[width=7cm]{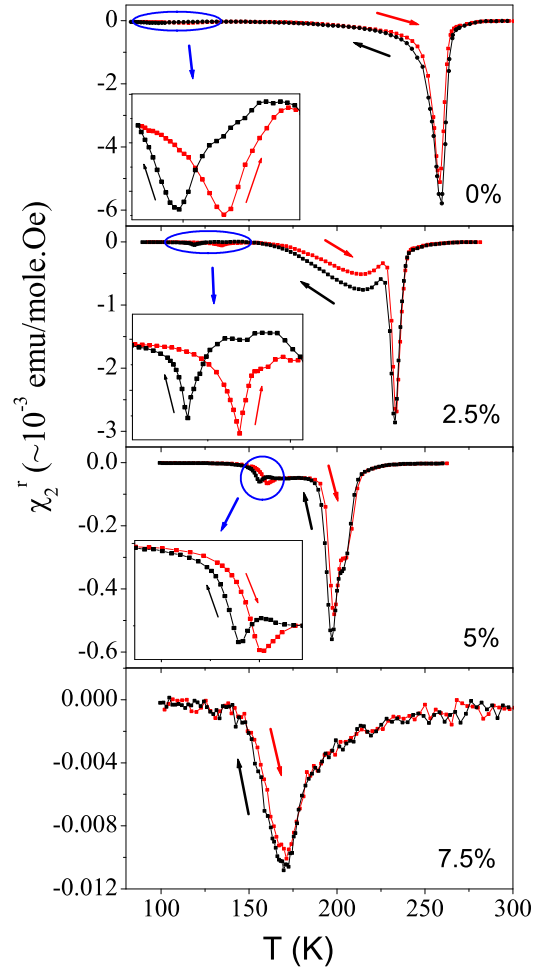}
		\caption{Temperature variation of the real part of second order susceptibility ($\chi_2^R$) measured during heating and cooling in an ac field of 9 Oe and frequency 131 Hz for the series $Pr_{0.5}Sr_{0.5}Mn_{1-x}Ga_xO_3$. The arrow shows the direction of temperature cycle. Insets show the magnified view of the same plots around the broad first order FM-AF transition which clearly depict the thermal hysteresis associated with the first order transition due to supercooling and superheating of FM/AF phases.}
	\label{fig:fig6}
\end{figure*}

\begin{figure*}
	\centering
		\includegraphics[width=7cm]{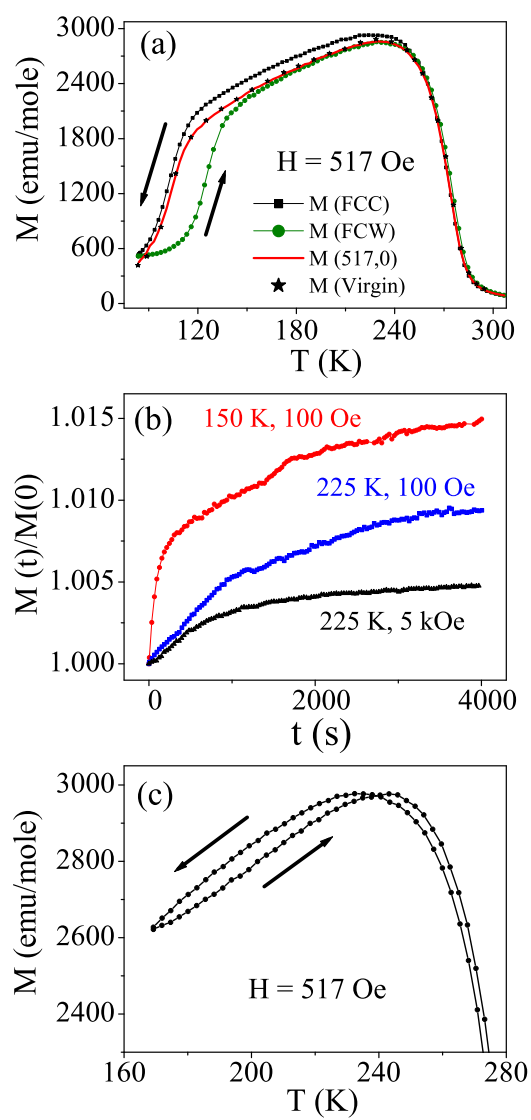}
		\caption{(a) DC Magnetization as a function of temperature has been measured in 517 Oe field for x = 0 compound. Measurements were done in FCC, FCW, M(517,0) and M(virgin) mode (defined in text). (b) Time (t) dependence of normalized magnetization has been plotted at different temperature and applied field. (c) This plot shows FCC and FCW magnetization measured in the same field but cooling was done only down to 170 K (much above $T_N$), which show reasonable TH in the FM state.}
	\label{fig:fig7}
\end{figure*}

\begin{figure*}
	\centering
		\includegraphics[width=7cm]{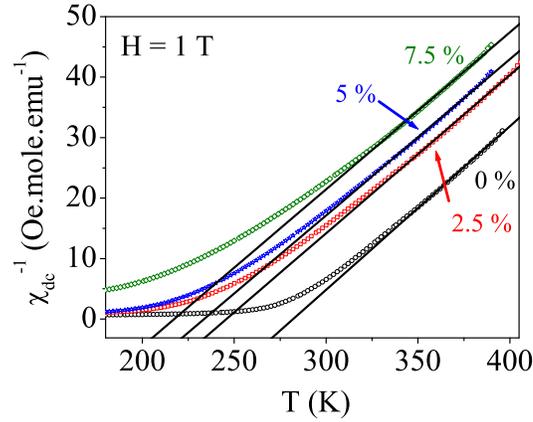}
		\caption{Inverse dc susceptibility measured in 1 T magnetic field has been plotted as a function of temperature for the series $Pr_{0.5}Sr_{0.5}Mn_{1-x}Ga_xO_3$. Straight lines show the Curie-Weiss law fitting of susceptibility data above $T_C$. Data for x = 0.025 compound has been shifted to higher temperature by 10 K for clarity.}
	\label{fig:fig8}
\end{figure*}

\begin{figure*}
	\centering
		\includegraphics[width=7cm]{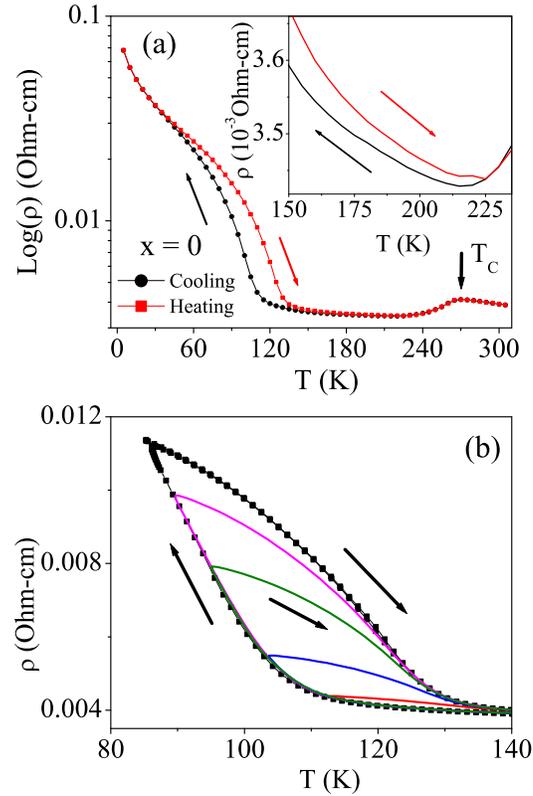}
		\caption{(a) The semi-log plot of resistivity as a function of temperature for x = 0 compound measured during heating and cooling. The arrow indicates the direction of temperature cycle. Inset shows the magnified view of thermal hysteresis in resistivity. (b) Minor hysteresis loop (MHL) around FM-AF phase transition has been plotted as a function of temperature. This plot shows disorder broadened first order transition. The lines inside the envelop curve represent the data recored in heating cycle.}
	\label{fig:fig9}
\end{figure*}

\begin{figure*}
	\centering
		\includegraphics[width=7cm]{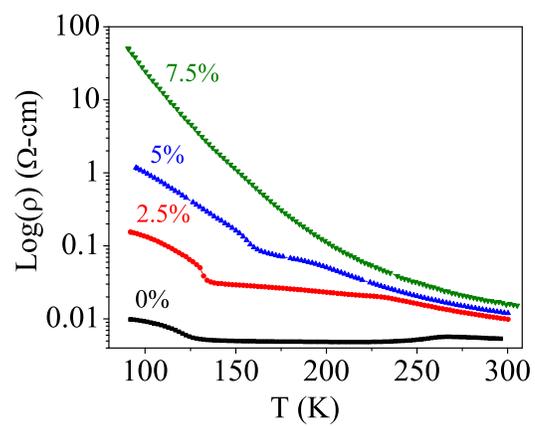}
		\caption{Resistivity data has been presented for the $Pr_{0.5}Sr_{0.5}Mn_{1-x}Ga_xO_3$ series as a function of temperature in a semi-log plot.}
	\label{fig:fig10}
\end{figure*}

\end{document}